\documentclass[onecolumn]{aastex6}

\usepackage{amsmath,amsfonts,amsthm}
\usepackage{amssymb}
\usepackage[T1]{fontenc}
\usepackage{graphicx}

\begin{document}
\title{The Brightest Galaxies at Cosmic Dawn from Scatter in the Galaxy Luminosity versus Halo Mass Relation}
\author{Keven Ren\altaffilmark{1,2}, Michele Trenti\altaffilmark{1,2} and Charlotte A. Mason\altaffilmark{3,4}}
\affil{$^1$ School of Physics, The University of Melbourne, Parkville, Victoria, Australia \\
$^2$ ARC Centre of Excellence for All Sky Astrophysics in 3 Dimensions (ASTRO 3D) \\
$^3$ Center for Astrophysics | Harvard \& Smithsonian, 60 Garden St, Cambridge, MA, 02138, USA\\
$^4$ Hubble Fellow\\}
\email{kevenr@student.unimelb.edu.au}


\begin{abstract}
The Ultraviolet Luminosity Function (UVLF) is a key observable for understanding galaxy formation from cosmic dawn. There has been considerable debate on whether Schechter-like LFs (characterized by an exponential drop-off at the bright end) that well describe the LF in our local Universe are also a sufficient description of the LF at high redshifts ($z>6$). We model the UVLF over cosmic history with a semi-empirical framework and include a log-normal scatter, $\Sigma$, in galaxy luminosities with a conditional luminosity function approach. We show that stochasticity induces a flattening or a feedback scale in the median galaxy luminosity versus halo mass relation, $L_{c}(M_{h})$ to account for the increase of bright objects placed in lower mass halos. We observe a natural broadening in the bright-end exponential segment of the UVLF for $z>6$ if processes that regulate star-formation acts on the same mass scale as at $z\sim5$, where the degree of broadening is enhanced for larger $\Sigma$. Alternatively, if the bright-end feedback is triggered at a near-constant luminosity threshold,  the feedback threshold occurs at progressively lower halo masses with increasing redshift, due to galaxies being more luminous on average at a fixed halo mass from rapid halo assembly. Such feedback results in a LF shape with a bright-end closer to that of a Schechter function. We include predictions for the $z>8$ UVLFs from future all-sky surveys such as WFIRST which has the potential to both quantify the scatter and type of feedback, and provide insight behind the mechanisms that drive star formation in the early Universe.

\end{abstract}


\section{INTRODUCTION}  \label{sec:intro}
The ultraviolet galaxy luminosity function (UVLF) is an effective tool to shed light on the physical processes that drive galaxy formation and evolution across cosmic history. UV light is predominantly emitted by young, short-lived massive stars, and as such is an effective tracer of the star formation history of a galaxy \citep{1996madau}.  Multiple efforts have characterized the UVLF up to $z\sim10$ using the Hubble Space Telescope, through surveys such as the Hubble Ultradeep Field (HUDF, \citealt{Bouwens2010}), the Cosmic Assembly Near-infrared Deep Extragalactic Survey (CANDELS, \citealt{2011ApJS..197...35G}), the Brightest of Reionizing Galaxies (BoRG, \citealt{2011ApJ...727L..39T}), the Frontier Fields \citep{2017ApJ...837...97L} and the Reionization Lensing Cluster Survey (RELICS; \citealt{Salmon2018}) which all search for objects during this epoch (e.g. \citealt{Bradley2012}, \citealt{Schmidt2014},  \citealt{2014ApJ...793L..12Z,2015ApJ...810L..12Z}, \citealt{2015ApJ...804L..30O, 2016ApJ...819..129O}, \citealt{2016ApJ...827...76B}, \citealt{2017ApJ...835..113L}, \citealt{Ishigaki2018}, \citealt{Morishita2018}). However, due to the small field of view of Hubble's infrared camera, these surveys have not achieved the sky coverage needed to probe the full dynamic range of high redshift ($z > 6$) UVLFs, particularly leaving the bright end loosely constrained as the number counts of these objects are expected to fall off rapidly. As efforts grow to probe this regime in the UVLF, there is some tension as to whether the exponential dropoff from the typical Schechter shape (having a functional form of $\phi(L) \propto (L/L_{*})^{\alpha}\exp(L/L_{*})$, where $L_{*}$ is a characteristic luminosity) in UVLF is preserved at the earliest times \citep{2015ApJ...803...34B, Stefanon2017}, or if a functional change, usually a double power law is a better fit the observation data (\citealt{2014MNRAS.440.2810B,Bowler2016}, \citealt{Ono2017}). 

Physical intuition suggests that the brighter the galaxy the more massive is its host halo, at least to first approximation. The technique of abundance matching encapsulates this idea, directly matching the cumulative UVLF with the underlying halo mass function (HMF) through the assumption that the relationship between halo mass and the corresponding galaxy luminosity is monotonic. This technique has been successful in describing the relation between stellar mass and halo mass at redshift $z\lesssim 4$ (e.g., see \citealt{Behroozi2010, Behroozi2013, 2018arXiv181205733A}). It also offers a natural opportunity to calibrate the relation between star formation and halo mass at a single redshift, and then construct minimal, yet effective semi-empirical models to predict the redshift evolution of the UVLF across time that take into account both the evolution of the HMF and of the characteristic halo/galaxy assembly time \citep{2010ApJ...714L.202T, 2013ApJ...768L..37T, 2015ApJ...813...21M, Mashian2015, Behroozi2015, 2018ApJ...856...81R}. Moreover, this method minimizes the number of assumptions needed to link star formation with the growth of its underlying dark matter halos. In fact, all complex interactions of baryonic physics in a dark matter halo that set the specific star formation efficiency are directly embedded in the calibration process. Despite its inherent simplicity, these models show remarkable predictive power to describe the evolution of the UVLF into the Epoch of Reionization, and are a competitive alternative to traditional numerical simulations that describe more of the physics involved at the expense of increased computational cost.

However, current high-z observations are primarily identifying $L\lesssim L_*$ galaxies, which are relatively common, and thus it is not so surprising that models based on average mass to light relations are adequate. In contrast, the role of stochasticity is expected to become increasingly prominent for the most luminous and rare sources. As the number density of the host halos that accommodate these sources grow exponentially from the massive end to lower masses, scatter in galaxy luminosity makes it possible for one of the more numerous smaller mass halo to host an over-luminous source, possibly altering the LF shape \citep{2005ApJ...627L..89C, Moster2010, Mashian2015}. Such variations in galaxy luminosity at fixed halo mass can be explicitly modeled under the conditional luminosity function approach (CLF; e.g. \citealt{Yang2003,Vale2004, 2005ApJ...627L..89C, Stanek2006}). The CLF method introduces a scatter parameter in the relation between galaxy luminosity and halo mass. This is in general a free parameter, although a minimum value for the scatter can be inferred from the probability distribution of the halo assembly times \citep{2018ApJ...856...81R}. 

In this paper, we extend previous semi-empirical modeling of the UVLF evolution combining it with a CLF model to account for the galaxy luminosity to halo mass scatter. We investigate predictions for the bright-end of the UVLF at $z>6$ and show that in presence of scatter, the median luminosity to halo mass relation needs to flatten at the bright-end to still be capable of reproducing the observed LFs. Furthermore, as a consequence of including scatter, we find that the most luminous galaxies are not in the most massive halos of the volume probed by a survey, but rather extreme outliers of over-luminous sources hosted in more common, lower mass halos. Finally, fixing the flattening at a constant luminosity leads to the expectation of the UVLF better preserving the exponential drop of a Schechter form at all redshifts, while if the flattening is fixed at a given halo mass, then the brightest end of the UVLF begins to broaden with redshift which can be possibly misinterpreted as a double power law. Future galaxy surveys combining data from the Wide-Field Infrared Survey Telescope (WFIRST, \citealt{2015arXiv150303757S}) and from the James Webb Space Telescope (JWST, \citealt{2006SSRv..123..485G}) will be able to discriminate between two scenarios, which in turn can be interpreted as different star formation feedback processes. 

This paper is structured as follows. Section~\ref{sec:model}, describes the model. Section~\ref{sec:results} presents the predictions for the UVLF at varying amounts of scatter. Finally we conclude and discuss implications of stochasticity present in other high-redshift objects in Section~\ref{sec:conclus}. Throughout this paper, we use the WMAP7 \citep{2011ApJS..192...18K} cosmological parameters with $\Omega_{m} = 0.272, \Omega_{b} =0.0455, \Omega_{\Lambda}=0.728, h=0.704, \sigma_{8}=0.81, n_{s}=0.967$. We use the \citet{2001MNRAS.321..372J} halo mass function. Magnitudes are given in the AB system \citep{1983ApJ...266..713O}.

\section{Modeling}  \label{sec:model} 
The model in this paper is based on a previous work, \citet{2018ApJ...856...81R}, extending the semi-empirical models developed by \citet{2010ApJ...714L.202T,2015ApJ...802..103T}, \citet{2013ApJ...768L..37T} and \cite{2015ApJ...813...21M} through inclusion of stochasticity in the galaxy luminosity versus halo mass relation, $L(M_h)$. The model foremost assumes that galaxy growth and formation is primarily driven through the assembly of the host dark matter halos. As such, we assume that the galactic star formation rate (SFR) is inversely proportional to the halo assembly period, defined as the time required to grow a halo of mass $M_{h}$ from an initial mass of $M_{h}/2$. Second, the model assumes that the stellar formation efficiency, i.e. the ratio of the host halo's mass to its stellar content, is only dependent on the halo mass and not redshift. The main consequence is the luminosity of a galaxy inside some halo of fixed mass is predominantly tied to the age of the stellar population, where higher redshift galaxies tend to host younger populations on average due to rapid assembly. With these assumptions and given a stellar population model, we can empirically calibrate the model at a single redshift with the ultraviolet galaxy luminosity function (UVLF) and a halo mass function (HMF) to determine the redshift independent stellar efficiency parameter. The evolution of the UVLF is subsequently governed by the evolution of the HMF and the properties of the synthesised stellar population in our galaxies. For this work, we will use the \citet{2001MNRAS.321..372J} HMF, and galaxies will be populated with the \citet{2003BRUZUAL&CHARLOT} simple stellar population (SSP) with a Salpeter initial mass function between $0.1M_{\odot}$ and $100M_{\odot}$ and constant metallicity $Z = 0.02Z_{\odot}$, appropriate for typical high-$z$ sources. We note that our choice of parameters for our SSP, and in particular the assumption of constant (low) metallicity, are expected to only marginally affect the evolution of the UVLF in the redshift range of interest for this study, at a level below or at most comparable to observational uncertainties in the LF determination. In fact, the calibration step removes first-order differences and typical galaxies at $z\gtrsim 5$ are expected to have low metallicities. 
One further basic assumption used to simplify modeling is that each halo is occupied by a single central galaxy, which is justified both by analytical halo occupation models and by analysis of cosmological simulations, which point to a $\lesssim 2\%$ contribution from satellites to the the UVLF at $z\gtrsim 6$ \citep{2018MNRAS.480.3177B}. 

\subsection{Conditional Luminosity Function}

\begin{figure}[ht!]
	\centerline{\includegraphics[angle=-00, scale=0.64]{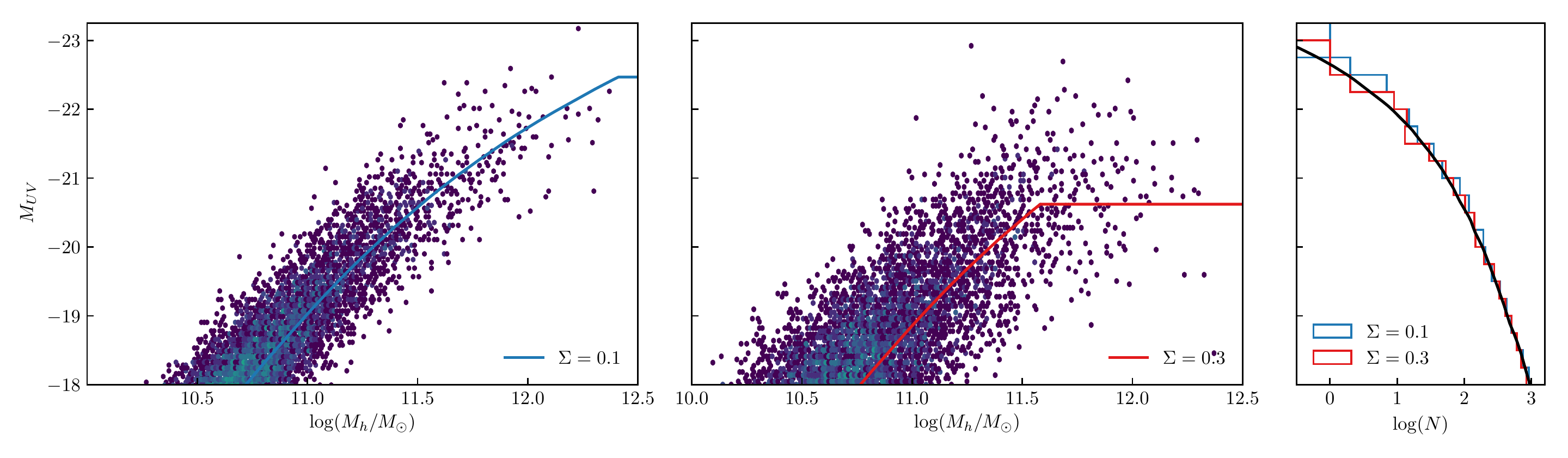}}
	\caption{\small Left and middle panels: Sampling of galaxy UV luminosities with the conditional luminosity function from the $z \sim 5$ halo mass function. Galaxies are stochastically assigned a luminosity based on the median galaxy luminosity to halo mass relation, $L_{c}(M_{h}, \Sigma)$ (solid lines) with a lognormal dispersion, $\Sigma = 0.1$ (left) and $\Sigma = 0.3$ (centre). We bin the total number of galaxies by magnitude (right). We note that our treatment of $L_{c}(M_{h}, \Sigma)$ preserves the luminosity function within this magnitude range. The $\Sigma = 0$ case is represented by the solid black line.}
	\label{fig:SBS}
\end{figure}

To first order, assuming an average $L(M_h)$ demonstrates remarkable power to describe and predict current observations \citep{2015ApJ...813...21M}. However, the data do not extend deep into the bright end ($L > L^{*}$) of the UVLF for high redshifts ($z \gtrsim 6$). This bright end of the UVLF is expected to show sensitivity to stochasticity in the luminosity versus halo mass relation because the number density of dark matter halos is functionally exponential at the massive end. Hence, to generally account for stochasticity in our modeling, we adopt a conditional luminosity function (CLF) approach to derive the UVLF. The CLF, $\Phi(\log L \mid M_{h})$, can be interpreted as the probability distribution of galaxy luminosities, $L$, from the median luminosity, $L_{c}$, given the host halo mass, $M_{h}$, 

\begin{equation}
\Phi(\log L \mid M_h)=\dfrac{1}{\sqrt{2\pi}\Sigma}\exp{\bigg( \dfrac{-\Big[ \log L - \log L_{c}(M_{h}, \Sigma, z) \Big]^{2}}{2\Sigma^{2}} \bigg) },
\label{eqn:clf}
\end{equation}

The log-normal dispersion $\Sigma$ is a free parameter in our modeling. Historically, it was introduced to best explain the scatter in the Tully-Fisher relation \citep{2005MNRAS.358..217Y}. When no scatter and redshift independence in stellar efficiency are assumed, the galaxies populate halos hierarchically over halo mass and thus the evolution of the LF would mirror the evolution of the underlying halo mass function. However, when scatter in galaxy luminosities is accounted, then there is a finite probability that a smaller halo can host a more luminous galaxy (Fig~\ref{fig:SBS}). This has a non-negligible effect on the shape of our overall LF, generally leading to an increase of the probability of observing $L>L^{*}$ galaxies for higher values of $\Sigma$. This increase in the number density of bright sources due to the exponentially increasing number of smaller halos from the massive end together with the log-normal scatter can be corrected to first order by applying a constant rescaling of the median luminosity to halo mass relation at the calibration redshift (see \citealt{2018ApJ...856...81R} for details).

To construct the usual luminosity function from the CLF, we integrate over the number density of dark matter halos, $\frac{dn}{dM_{h}}$ weighted by the probability a halo $M_{h}$ can host a galaxy of luminosity, $L$ given a dispersion $\Sigma$, 

\begin{equation}
\phi(\log L)
 = \int_{0}^{\infty} \dfrac{dn}{dM_{h}} \Phi(\log L \mid M_{h})dM_{h},
\label{eqn:lf}
\end{equation}

For the $\Sigma = 0$ case, we follow the model of \citet{2015ApJ...813...21M}, adjusting for cosmology and HMF. We calibrate our model to the \citet{2015ApJ...803...34B} $z \sim 5$ UVLF. The evolved $z > 5$ LFs show good agreement with observations and have well-described Schechter functions associated with them. We then select a value of $\Sigma$ for Equation~\ref{eqn:lf} and subsequently determine the median halo mass-galaxy luminosity relation, $L_{c}$ that best recovers the LF at $z \sim 5$ (our calibration redshift).

\subsection{Methodology}

\begin{figure}[ht!]
	\centerline{\includegraphics[angle=-00, scale=0.84]{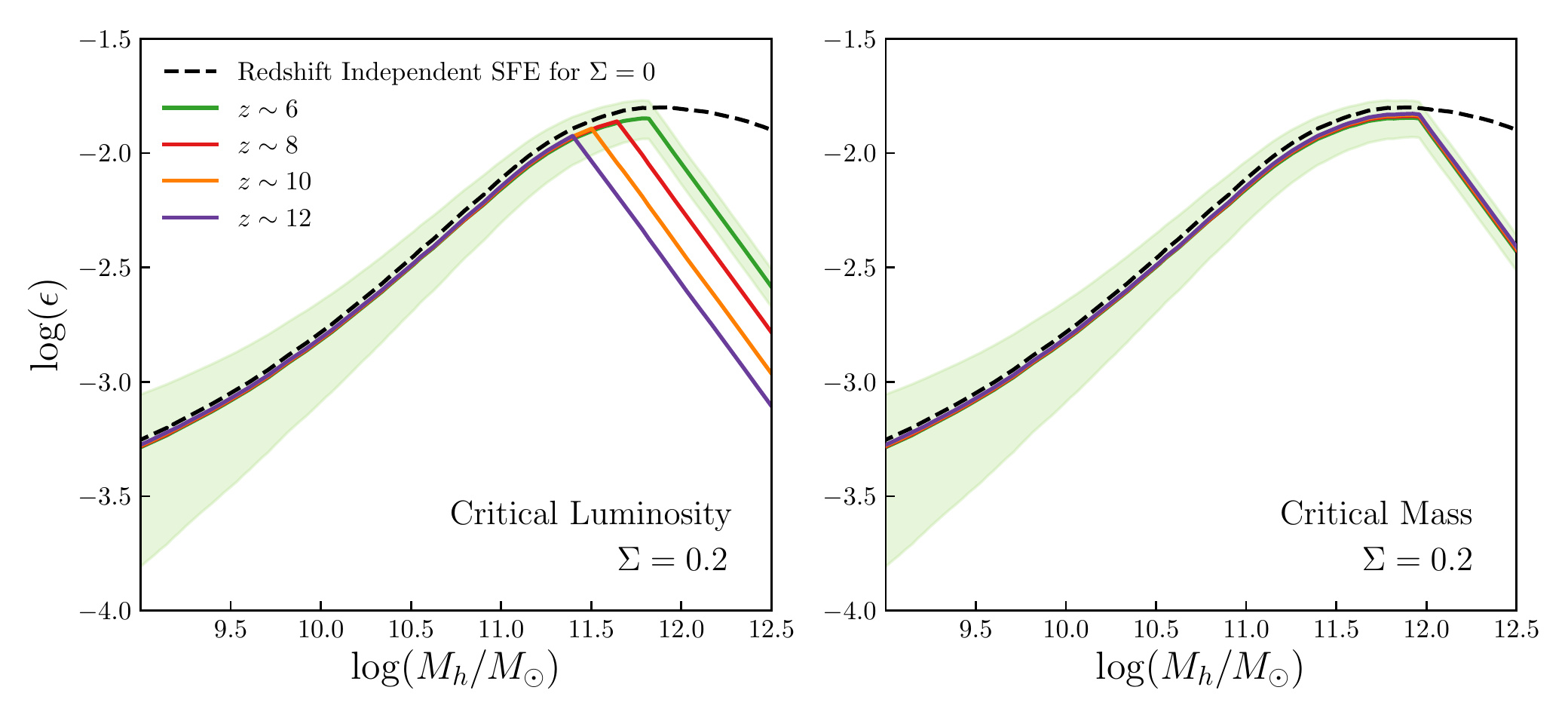}}
	\caption{\small Stellar efficiency, $\varepsilon = M_{*}/M_{h}$ as a function of halo mass, $M_{h}$. The solid lines are $\varepsilon$ where the UVLF evolves in accordance with a flattening from a critical mass threshold (left) or a critical luminosity threshold(right). The scatter modeled here is for $\Sigma = 0.2$. The dashed line is the redshift independent SFE from the $\Sigma = 0$ modeling. The typical $1\sigma$ uncertainty is included for solid green $z\sim 6$ line.}
	\label{fig:sfe}
\end{figure}

To derive $L_c(M_h,\Sigma,z)$ at the calibration redshift in the presence of scatter, one possibility is to use the iterative deconvolution technique described in \citet{Behroozi2010} to calculate $L_{c}$. This method has been used successfully in deriving $L_{c}$ analogues (e.g. \citealt{Reddick2013, Sun2016, 2018arXiv181205733A}), but its applicability to our study can be limited if $\Sigma$ is too large (i.e. in presence of a significant amount of scatter), because an exact solution to the deconvolution may not exist under these conditions. In addition, deconvolution algorithms may introduce unphysical oscillations in $L_{c}$ for the case of slow convergence when a large number of iterations is required (Appendix~\ref{apdx:a}). Therefore we take here the alternative approach to cast the derivation of $L_c(M_h,\Sigma,z)$ as an optimization problem. Appendix~\ref{apdx:a} shows the results from iterative deconvolution and compares them against our optimization approach. 

Thus, we employ a trust region algorithm to minimize the square of the residuals between our modeled LFs from Equation~\ref{eqn:lf} and the LF at $z \sim 5$, under the constraint that $\log L_{c}(M_{h})$ is a monotonically increasing function in $M_{h}$. By definition the $z\sim 5$ LF is equivalent to the \citet{2015ApJ...803...34B} $z \sim 5$ LF. The best fit relations for $L_{c}(M_{h},\Sigma > 0)$ can be obtained using the initial guess $\log L_{c}(M_{h}, \Sigma = 0)$, but achieving convergence is time consuming. Since the qualitative trend is that of a flattening beyond a characteristic luminosity/halo mass, we retain the spirit of a model as simple as possible, and thus further limit the degrees of freedom in the minimization to two parameters, i.e. a constant scaling in the luminosity (which is needed to describe the faint-end of the UVLF), and a flattening of $\log L_{c}(M_{h})$ above a critical mass/luminosity threshold (see Fig.~\ref{fig:SBS}).

To summarise, the steps to derive the median halo mass-galaxy luminosity relation given $\Sigma$ are:

\begin{enumerate}
  \item A log-normal dispersion in the median galaxy luminosity, $L_{c}(M_{h})$ increases the galaxy luminosity on average but preserves the power law slope in the LF \citep{2005ApJ...627L..89C}. We account for this with a constant scaling in the median luminosity, $L'_{c}(M_{h}) = L_{c}(M_{h})/k(\Sigma)$  (or equivalently, $\log L'_{c}(M_{h}) = \log L_{c}(M_{h}) - k(\Sigma)$), for a constant $k(\Sigma)$, minimizing the residuals between modeling and calibration LF for a power slope segment (taken to be $-18 < M_{UV} < -15$). Consequently, rescaling $L_{c}(M_{h})$ slightly reduces the stellar mass content per unit halo mass (Fig.~\ref{fig:sfe}) to compensate for the brighter average galaxy luminosities.

  \item For $M_h>M_{h}^{C}$, we require a flattening in the median halo luminosity to retain close correspondence to the calibration LF. For this, we flatten the relation above a critical galaxy luminosity, i.e. $L_{c}'(M_{h} > M_{h}^{C}) = L_{c}'(M_{h}^{C})$ and minimize the residuals between calibration LF and model in the magnitude range ($23.5 < m_{UV} < 27$). A flattening applied on $L_{c}(M_{h})$ implies a corresponding flattening in the stellar mass content for halos more massive than the threshold $M_{h}^{C}$. As such, the star formation efficiency in such halos would scale as $\sim 1/M_{h}$ (Fig.~\ref{fig:sfe}). From this, we can typically expect an upper limit to the stellar mass for the brightest galaxies, while the luminosity is approximately inversely proportional to the stellar population age.
  
\end{enumerate}

\subsection{Calibration and UVLF Evolution Options}

The critical point $L_{c}(M_{h}^{C})$ determined during calibration at $z\sim 5$ is used to predict the UVLF evolved in redshift. From this, we investigate two distinct scenarios at how the UVLF could evolve:

1) The scaled median halo mass to galaxy luminosity relation, $L'_{c}(M_{h},z) = k(\Sigma) \times L(M_{h},z)$ inherits a critical intrinsic luminosity where flattening is applied, i.e, we set $L'_{c}(M_{h},z) = L'_{c}(M_{h}^{C},z)$ when $L'_{c}(M_{h},z) > L'_{c}(M_{h}^{C},z)$. In this scenario, the exponential slopes of the calibration LF are approximately preserved, particularly so when $\Sigma \lesssim 0.2$. However, the star formation efficiency is no longer redshift independent as the halo mass for the cut-off decreases as $z$ increases (see Fig.~\ref{fig:sfe}). Note that the effects of dust absorption are included in the model and handled self-consistency to observations. Specifically, a dust correction is applied before calculating the observed luminosity at each redshift. The dust-correction method we adopt is the method used in \citet{2015ApJ...802..103T} and \citet{2015ApJ...813...21M} in conjunction with the observations of the UV continuum slopes from \citet{2015ApJ...803...34B}.

2) The scaled median halo mass to galaxy luminosity relation, $L'_{c}(M_{h},z) = k(\Sigma) \times L(M_{h},z)$ inherits a critical halo mass where flattening is applied for subsequent redshifts, i.e, we set $L'_{c}(M_{h} > M_{h}^{C},z) = L'_{c}(M_{h}^{C},z)$. Here, the bright end of the LF is broadened at higher redshift, with an increasing effect for larger values of $\Sigma$. In this case, we maintain the redshift-independence on the stellar efficiency parameter (see Fig.~\ref{fig:sfe}).

For either scenarios, we consider a domain of $\Sigma$ to probe. We note that the parameter $\Sigma$ is not fully constrained. From our modeling, the first order stochasticity term comes from the distribution in halo assembly times with an estimated lower limit of $\Sigma \sim 0.2$ for $z > 2$ \citep{2018ApJ...856...81R}. While the stochasticity in halo assembly times is effectively redshift independent at high $z$, the value does not account for other galaxy formation processes, such as galaxy-to-galaxy variations in dust extinction and/or stellar efficiency. In turn, these additional contributions may be redshift dependent and boost the effective scatter. To derive an estimate on the possible value of $\Sigma$, we turn to outputs from semi-analytical modeling. A snapshot at $z\sim 8$ from the Meraxes simulation \citep{2016MNRAS.462..250M} gives a value $\Sigma \sim 0.4$ when considering halos in the mass range of $M_{h} > 10^{10} M_{\odot}$. This value is not a formal upper limit of $\Sigma$, but rather is used as a guidance for the investigation on the impact of scatter between the minimal case where $\Sigma \sim 0.2$ and another possible regime at $\Sigma \sim 0.4$. Thus, here we consider a range $0.2 < \Sigma < 0.4$ to investigate the effects on the UVLF and on the galaxy number count forecasts.

\subsection{Line of Sight Gravitational Magnification}\label{sec:lensing}

We also include the distortion of the UVLF due to gravitational lensing along the line-of-sight. Due to intervening structure high-redshift galaxies can be gravitationally lensed which causes a `pile-up' of magnified sources at the bright-end of the LF, and the probability of being lensed increases as the source redshift increases \citep{Wyithe2011,2015MNRAS.450.1224B,2015ApJ...805...79M}. \citet{2015ApJ...805...79M} demonstrated that this magnification bias dominates at $M_{UV} \lesssim -24$, and so its effects should be included when considering the brightest end of the UVLF \citep{Ono2017}.

We follow the methods of \citet{2015ApJ...805...79M} to calculate lensed LFs in this work and refer the reader there for more details. The probability of a line-of-sight intersecting a lens, the lensing optical depth, $\tau_m$, is calculated as a function of redshift for an evolving lens population. Lenses are modeled as singular isothermal spheres (SIS), with Einstein radii estimated using a redshift-dependent velocity dispersion function. In this work we consider only strong lensing magnification $\mu \geq 2$, i.e. only the brighter image of an SIS lens, as magnifications $\mu < 2$ have negligible impact on the bright end LF \citep{2015ApJ...805...79M}.

For a UVLF $\phi(L)$ the lensed LF is given by:

\begin{equation}
\label{eqn:lf_lensed}
\phi_{lensed}(L) = (1 - \tau_m) \frac{1}{\mu_{demag}}\phi\left(\frac{L}{\mu_{demag}}\right) + \tau_m \int_2^\infty d\mu \, \frac{p(\mu)}{\mu} \phi\left(\frac{L}{\mu}\right)
\end{equation}

where the first term ensures the mean magnification on the whole sky is one, with $\mu_{demag} = \frac{1 - \tau_m \langle\mu\rangle}{1 - \tau_m}$. $p(\mu) = 2/(\mu-1)^3$ is the magnification distribution for an SIS lens and $\langle\mu\rangle = 3$ is the mean magnification of this distribution.

\section{Results and Discussion} \label{sec:results}

\begin{figure}[ht!]
	\centerline{\includegraphics[angle=-00, scale=0.84]{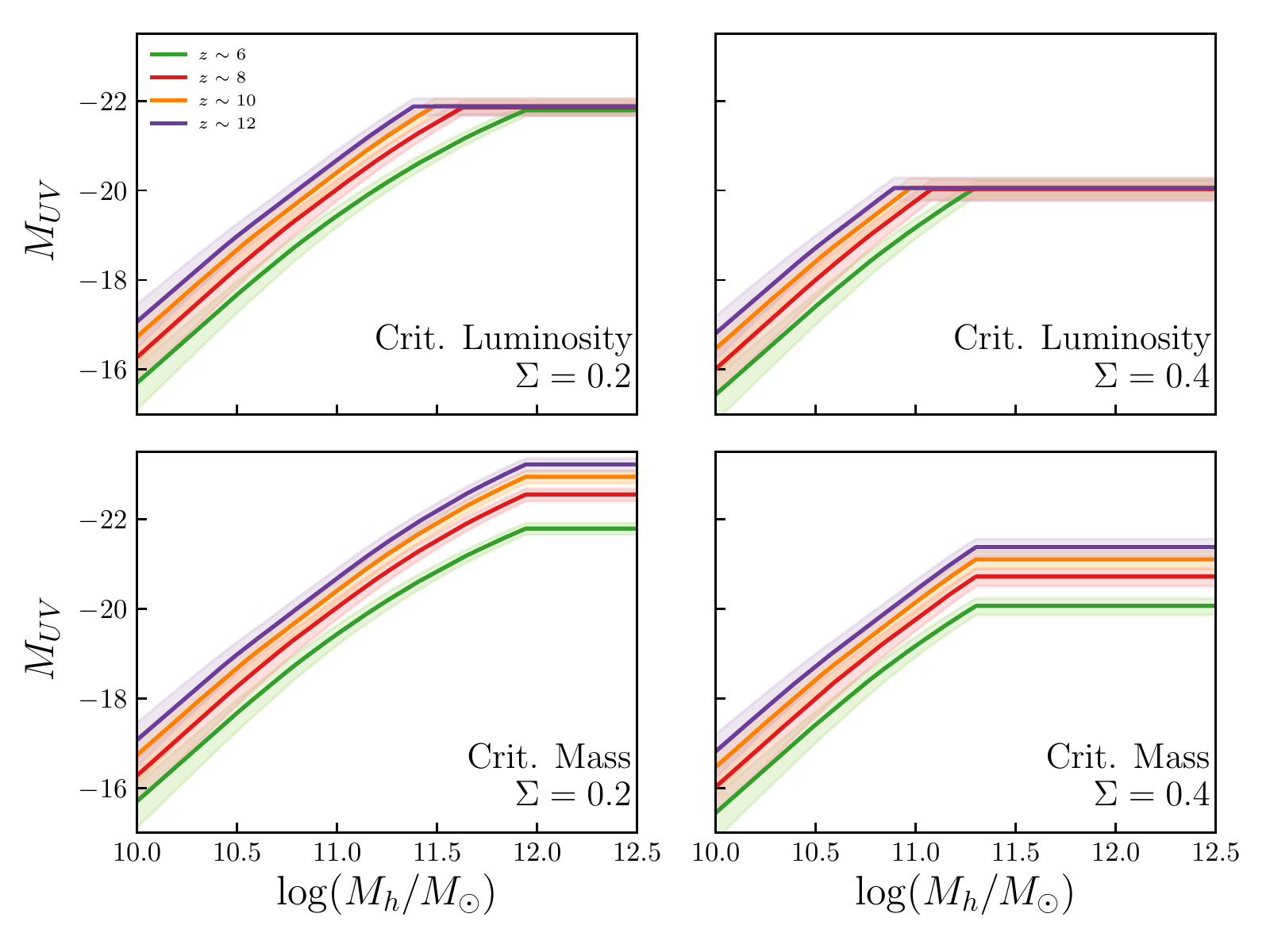}}
	\caption{\small The median halo mass to UV galaxy luminosity relation, $L_{c}(M_{h})$ at different redshifts, $z \sim 6,8,10,12$ (colored lines) after dust-correction for $\Sigma = 0.2$ (left) and $\Sigma = 0.4$ (right). The (upper) plotted scenario assumes that the flattening threshold is at constant luminosity and redshift independent, while the (lower) plotting cases assumes that the flattening threshold is constant mass and redshift independent. The shaded regions are the $1\sigma$ uncertainties.}
	\label{fig:LvM}
\end{figure}

\begin{figure}[ht!]
	\centerline{\includegraphics[angle=-00, scale=0.84]{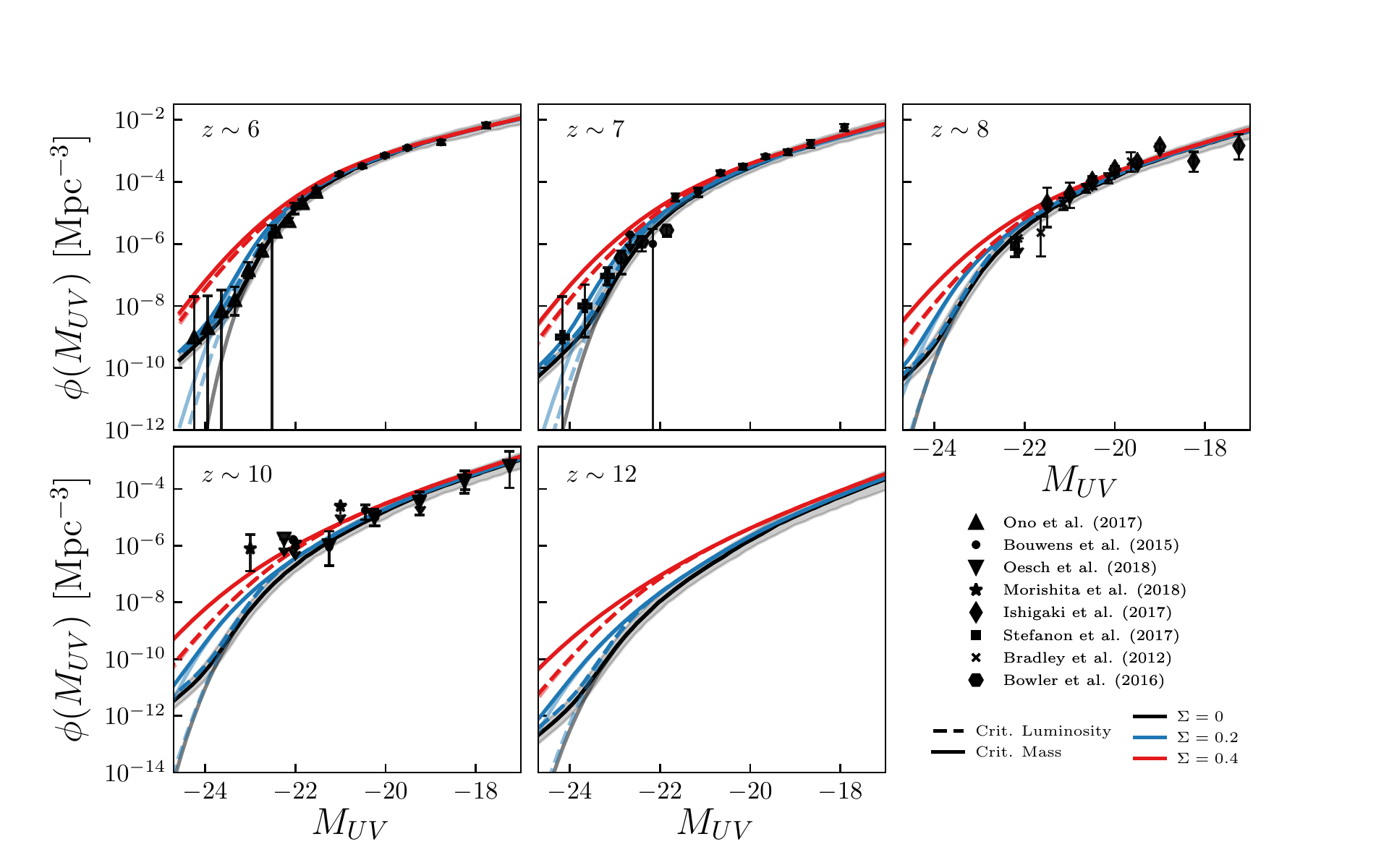}}
	\caption{\small The modeled ultraviolet luminosity functions with log-normal scatter values of $\Sigma = 0,0.2,0.4$ (colored lines) at various high redshifts: $z\sim 6$ (upper left), $z\sim 7$ (upper center), $z\sim 8$ (upper right), $z \sim10$ (lower left), $z\sim 12$ (lower center). The line styles corresponds to whether the median halo mass to galaxy luminosity relation has a constant luminosity cut-off (dashed) or a constant mass cut-off (solid). The unlensed LF are underlaid with slight transparency. The typical 1$\sigma$ uncertainty is shown for reference in the $\Sigma = 0$ curve. Recent data points from observations are shown as solid black points.}
	\label{fig:LFs0}
\end{figure} 

\begin{figure}[ht!]
	\centerline{\includegraphics[angle=-00, scale=0.74]{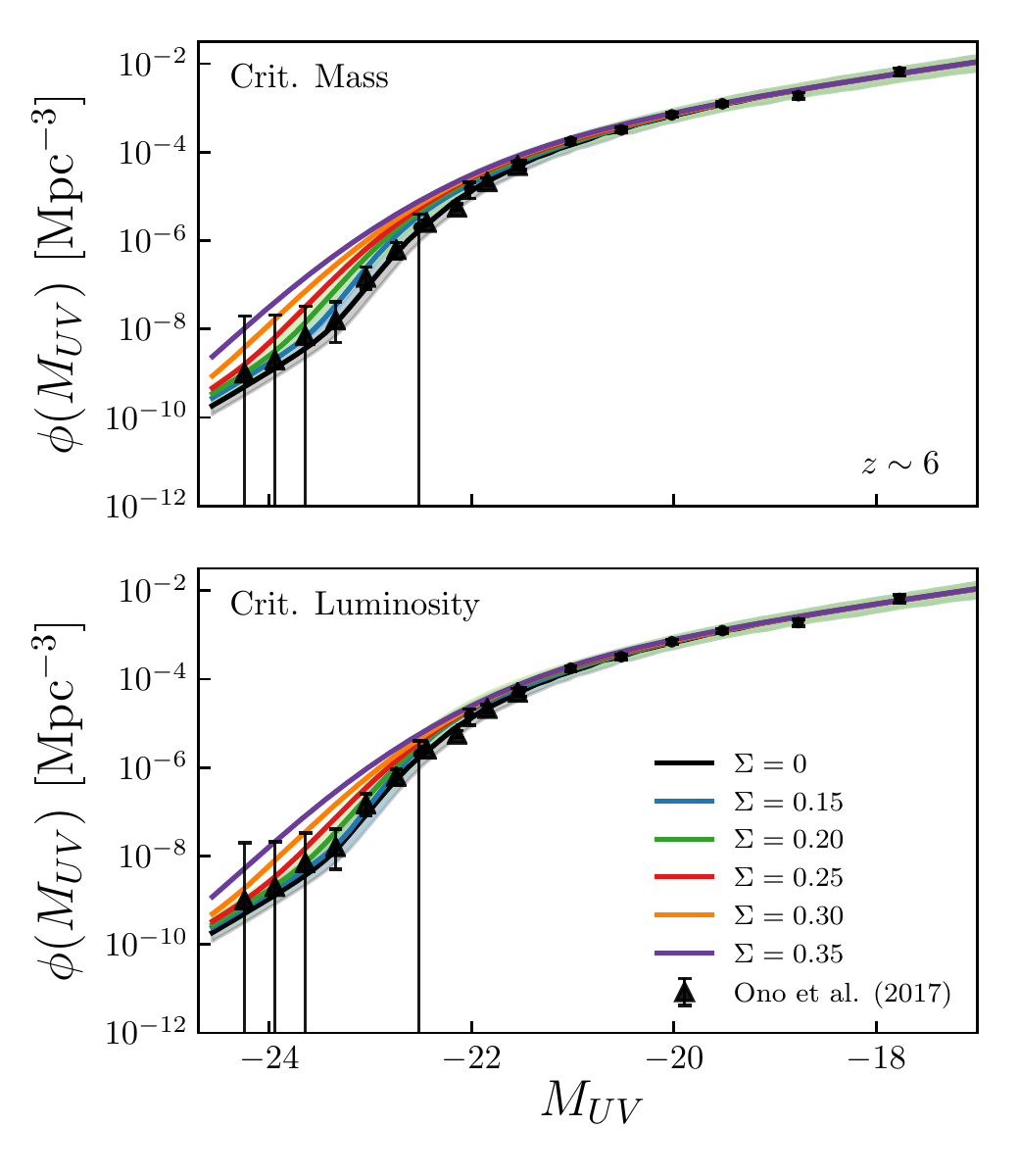}}
	\caption{\small Evolved $z\sim6$ luminosity functions assuming various values of $\Sigma$ and including line of sight gravitational lensing. We consider two scenarios when star formation is modulated by a: (upper) critical mass threshold and (lower) critical luminosity threshold. The black points are the observations. We include the 1$\sigma$ uncertainties for $\Sigma = 0, 0.15, 0.2$ for reference.}
	\label{fig:Sc5}
\end{figure} 

In Figure~\ref{fig:LvM}, we show the median halo mass to galaxy luminosity relations, $L_{c}(M_{h})$ for various $\Sigma$ and redshifts while assuming the scenarios of: (1) a critical luminosity threshold is applied (upper panels) and (2) a critical mass threshold is applied (lower panels). These characteristic values where flattening is applied are determined by the calibration of the UVLF at $z \sim 5$. Irrespective of the mechanism for flattening, higher values of $\Sigma$ require a lower mass/luminosity critical point, since larger $\Sigma$ give lower mass halos better odds to host a galaxy with a luminosity that is significantly greater than the median luminosity.

The resulting ultraviolet luminosity functions (UVLFs), inclusive of line-of-sight lensing magnification, are shown in Figure~\ref{fig:LFs0}. The black solid line is the $\Sigma = 0$ model and is characterised by its steep exponential drop-off at its bright end. In addition, the redshift evolution of the no-scatter case shows a good fit to all current observations. However, for a non-zero $\Sigma$ we begin to see marked differences in the evolution for the UVLF. 

For the critical mass model (solid colored lines), the UVLF experiences a broadening in the shape of the exponential end, which becomes increasingly prominent at higher redshifts. This is because the characteristic mass in the halo mass function (defined as the knee point where the transition from exponential to power law number density occurs) decreases with increasing redshift. Thus, at higher redshift there is a progressively larger impact of over-luminous galaxies hosted in lower mass halos if the $L_c(M_h)$ curve is flattened at constant $M_h$.

On the contrary, if we assume the scenario for a critical luminosity threshold (dashed colored lines), we see that the UVLF has a steeper bright end compared to the critical mass case. A critical luminosity threshold effectively reduces the mass threshold to apply flattening at increasing redshift. This is because halos at fixed mass typically have a more rapid assembly at higher redshifts and the galaxies they host find it easier to reach critical luminosity threshold due to their younger stellar populations. For a scatter $\Sigma = 0.2$, we find that the calibrated luminosity of $M_{UV} \sim -22$ is very similar to the UVLF predictions obtained for the no-scatter model with $\Sigma = 0$. The comparatively larger number of halos in this `flattened' regime is sufficient to otherwise offset the broadening from scatter in the bright end of the UVLF. In fact, for all $\Sigma$ values $\lesssim 0.2$, the UVLF is consistent with the $\Sigma = 0$ model throughout cosmic evolution. One caveat for $\Sigma > 0.2$ cases is that the bright end of the UVLF can significantly depart from a Schechter-like function despite calibration. This occurs simply because it becomes impossible to recover the steep exponential bright end in the observed LF with the forward application of Equation~\ref{eqn:lf} when there is a sufficiently broad distribution of $\Sigma$. Since $\Sigma$ is the dominant parameter in shaping the LF's bright end through the sampling of over-luminous objects from the numerous lower mass sources, there is a limit on $\Sigma$ if one wants to retain consistency with $z\sim6$ observations. 

Here, we exploit the discrepancies in the $z \sim 6$ UVLF for higher values of $\Sigma$ for a tentative measure of the upper limit of $\Sigma$ as observations suggest a steep bright end. We show this in Figure~\ref{fig:Sc5} for the evolved $z\sim6$ LFs with a range of $\Sigma$ values. The over-broadening of the LF from scatter points to an upper limit of $\Sigma \lesssim (0.25) 0.3$ for the critical mass (critical luminosity) cases and assuming our particular model of gravitational lensing. From this, we realise a slight tension between the possible values of $\Sigma$ inferred from this study and that from the Meraxes semi-analytical simulations. However, we attribute this to the smaller volume of Meraxes (with box length $\sim 100$Mpc), which implies that the simulation galaxy formation model might have been tuned to more typical objects, making in turn more challenging to address with high confidence the properties of extremely rare galaxies. 

From the plots in Figures~\ref{fig:LFs0} and~\ref{fig:Sc5}, we note that since the brightest end of the observed LF cannot be recovered exactly, we should also expect a degeneracy in the two scenarios for $\Sigma > 0.2$, where a lower-$\Sigma$ critical mass threshold case can share a $z > 6$ LF curve similar to that of a higher-$\Sigma$ case assuming a critical luminosity threshold. Discrimination between the two scenarios at a given redshift is challenging using only the LF, and would require an additional independent measure of $\Sigma$ such as from the local clustering strength, which can be obtained either from the two point correlation function or from number counts of neighbour galaxies. 

The flattening we imposed on $L_{c}(M_{h})$ can be interpreted as a feedback process that sets in past a certain critical point. How this relation evolves in redshift depends on the dominant feedback processes that prevent gas infall to form new stars. In one scenario, the mechanism behind having a critical mass at which star formation is suppressed could be attributed to radio-mode active galactic nuclei (AGN) feedback. Phenomenologically, star formation is truncated when the energy output from the accretion of hot gas in the halo onto the central supermassive black hole is sufficient to prevent gas from the disk from cooling further \citep{2006MNRAS.365...11C}. This type of feedback is dependent on the host halo mass and is dominant when the halo is sufficiently massive enough to contain a large amount of quasi-static hot gas. Conversely, we could also postulate a scenario where $L_{c}(M_{h})$ evolves with some hard threshold based on AGN luminosity, which is in turn linked to galaxy luminosity. Here, a possible mechanism could be a regulation based on the rate of halo growth, e.g. the host galaxy luminosity could be limited by quasar activity. If a halo assembles too rapidly, i.e. the enhancement in the accretion from the cold gas disk into the supermassive black hole can be substantial enough to quench additional star formation in the halo. However in principle, both mechanisms, critical mass or luminosity could impact the halo mass-galaxy luminosity relation in combination, and suppress the total amount of broadening relative to a pure scenario for a critical mass threshold. Hence, to properly constrain $\Sigma$, observations of the bright end extending into rare objects would be required at multiple redshifts.

Current measurements for high redshift ($z > 6$) UVLFs populate the fainter power-slope section, and are at the cusp of the exponential $L > L^{*}$ region. Thus, we show in this work that reasonable amounts of scatter will effectively induce a broader exponential slope over the usual Schechter function used in conventional fitting practices. In this regard, we already see some tension from using the standard Schechter parameterization of the observed UVLF at high redshifts. For example, earlier observations of the broadening in the bright end galaxy UVLF is seen in \citet{Bowler2015} who suggests that the $z \sim 7$ UVLF can be better fit with a double power law. More recently, the Subaru's GOLDRUSH Hyper Suprime-Cam program probing $\sim 100$ deg$^{2}$ of sky also sees substantial broadening in the bright end and recommends either a double power law or if lensing is predominant, a lensed Schechter function to fit the $4 < z < 7$ UVLFs \citep{Ono2017}. In both of these cases where $\Sigma \lesssim 0.25$, we find that our lens-broadened UVLF is a good fit to observations in the redshift ranges $6 < z < 7$ without invoking a functional change. For $z \sim 8$, the observations have not reached far enough into the bright end of the UVLF to make a definite conclusion to the broadening of the UVLF. Likewise, in the very high redshift regime ($z > 10$) we see a similar scenario where observations have not probed far enough into the bright end. 

We should note that while $\Sigma \sim 0$ is perfectly consistent with existing observations to all redshifts considered, a scatter free scenario is not without tension. The observation of the unexpectedly bright ($M_{UV} = -22.1$), $z=11.1$ galaxy, GN-z11 \citep{Oesch2016} finds that the extrapolated Schechter parameterizations of the UVLF sees it unlikely to capture a galaxy of similar magnitude given the area probed, which may also point to a degree of broadening from $\Sigma$, at least at redshifts $z > 10$ to reconcile modeling with observations. More importantly, the cosmological impact of $\Sigma$ extends beyond modeling the bright end of the UVLF, as it affects the spatial distribution of galaxies and their clustering properties as well, in particular for halos hosting extreme objects. 

In presence of $\Sigma \gg 0$, lower mass halos are those preferentially hosting the brightest and rarest galaxies, and in turn we would thus expect $\Sigma$ to have some influence on local-scale clustering around brightest objects. Specifically, the larger $\Sigma$ is, the more reduces is clustering compared to standard modeling that assumes a vanilla abundance matching (plus duty cycle) \citep{2018ApJ...856...81R}.

The inclusion of stochasticity to the semi-empirical modeling of \citet{2010ApJ...714L.202T}, \citet{2013ApJ...768L..37T} and \cite{2015ApJ...813...21M} is its natural first-order extension to better describe the rarer luminous galaxies that reside at the earliest epochs of cosmic history. This extended model offers two pathways in the evolution of the median galaxy luminosity to halo mass relation each with different implications on the mechanisms that drive star formation, and each with differing impact on the eventual shape of the bright end of the UVLF. We note that the amount of broadening in the UVLF is dependent on the distribution of $\Sigma$. In reality, the amount of broadening could be enhanced. In our modeling $\Sigma$ was assumed lognormal based on observations of the Tully-Fisher relation \citep{2005MNRAS.358..217Y} and in agreement of assumptions used in prior modeling by \citet{2005ApJ...627L..89C, Vale2008, Munoz2011}. However, a lognormal distribution of $\Sigma$ may not necessarily be the case. For example, if the scatter is predominantly driven by the assembly of the dark matter halos then we would expect more broadening, as the asymmetry in the halo assembly times results to a slight skew in the scatter of galaxy luminosities towards the brighter end \citep{2018ApJ...856...81R}. Furthermore, contributions from other sources of scatter (for cases where $\Sigma > \Sigma_{\mathrm{minimum}}$) could also alter the distribution and thus the amount of broadening experienced by the exponential end of the UVLF. 

\begin{figure}[ht!]
	\centerline{\includegraphics[angle=-00, scale=0.64]{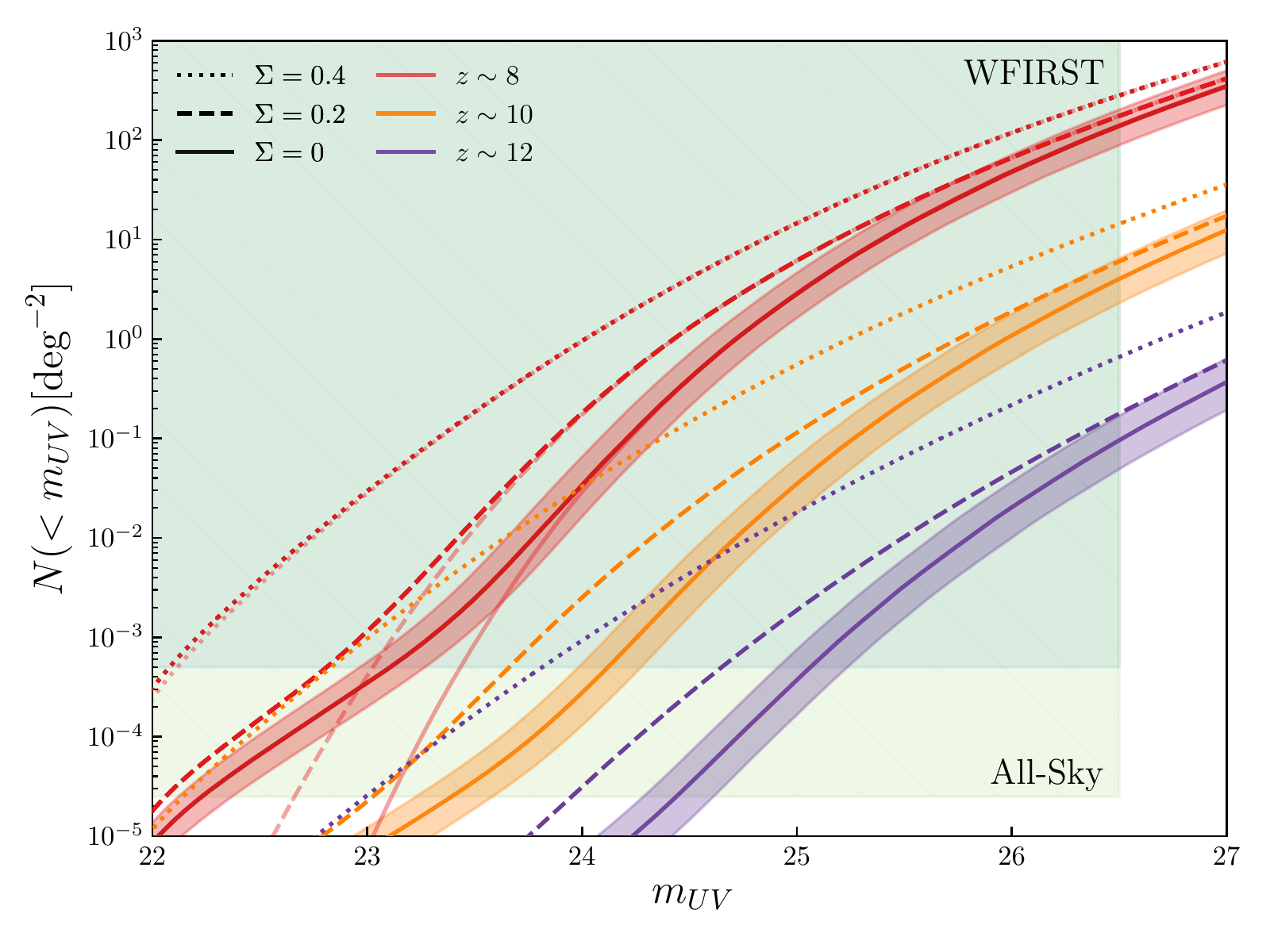}}
	\caption{\small Modeled cumulative number counts of galaxies with $< m_{UV}$ per square degree at different values of scatter $\Sigma = 0,0.2,0.4$ (solid, dashed, dotted respectively) and at different redshifts, $z \sim 8, 10, 12$ (colored lines). The effects on the UVLF from gravitational lensing along a line of sight is included. The unlensed, intrinsic number counts for $z\sim8$ curves are underlaid with slight transparency. Here we assume the scenario that the halo mass to galaxy luminosity relation contains a mass dependent, redshift independent feedback process. The shaded dark region is the WFIRST field of coverage of $\sim 2000 \mathrm{~deg}^{2}$ and the lighter region encompasses the all-sky coverage of $\sim 40000 \mathrm{~deg}^{2}$. The typical $1\sigma$ uncertainty is included for the $\Sigma=0$ cases.}
	\label{fig:WFIRST}
\end{figure}

To constrain the shape of the UVLF better, deriving limits on $\Sigma$ would be useful. One viable observational approach to measure/constrain $\Sigma$ is through shallow large area surveys looking for the brightest objects of each redshift. The first of such upcoming surveys, WFIRST will probe an effective volume down to $\sim 10^{10} \mathrm{Mpc}^{3}$ in the relevant observing range. In Figure~\ref{fig:WFIRST}, we assume the scenario of a critical mass threshold and we show the cumulative number of objects expected per square degree accessible by WFIRST ($\sim2000$ deg$^{2}$) and as well as the objects expected over the total all-sky area ($\sim 40000$ deg$^{2}$). As the brightest end of the high redshift UVLF can also be substantially populated by lensed sources, we also include the effects of line of sight gravitational lensing from intermediate structures (Section~\ref{sec:lensing}; see also \citealt{Wyithe2011}). This effect can become significant for small values of $\Sigma$ where the bright end of the LF is still steep. This is most visually prominent in the $z\sim 8$ curve for $0 \lesssim \Sigma \lesssim 0.2$ where we see a `kick' for sources $m_{UV} <23$ due to lensing and this mimics the impact of scatter. Because of this, we find a preferential window for counting $m_{UV} <23.5$ sources which still allows for discrimination between $\Sigma$ values. For $z\sim 8$ galaxies we expect a factor of $\gtrsim 10^{1}$ increase in the number of $m_{UV}< 23.5 $ galaxies even with minimal scatter ($\Sigma = 0.2$) compared to the ($\Sigma = 0$) modeling. These trends are further expected to be enhanced if $\Sigma$ is significantly higher than the lower limit estimate and at higher redshifts. Here, a value of $\Sigma = 0.4$ would see a factor $\gtrsim 10^{2}$ increase in $m_{UV}<23.5$ galaxy number counts at $z \sim 8$. However, any interpretation to the value of $\Sigma$ through this method could be made more complicated if the mechanism for a constant luminosity cut-off is significant at any scale, due to a degeneracy in mechanism for $\Sigma \gtrsim 0.2$. If the mechanism responsible for a critical luminosity threshold is fully dominant, then large scale surveys can be rendered an ineffective method of determining the value of $\Sigma$ (if the true value of $\Sigma \lesssim 0.2$) since the shape of the UVLF is no longer sensitive to $\Sigma$ in that regime (see Fig~\ref{fig:LFs0}).

To build a more robust measure for $\Sigma$, a combined measurement from large scale surveys supplemented with measures of $\Sigma$ from other galaxy properties, such as their clustering would be able to comprehensively discriminate between the two scenarios of critical luminosity or critical mass thresholds and shed light on actual value of $\Sigma$. 

\section{Conclusion and Acknowledgements} \label{sec:conclus}

In this work, we extend the simple semi-empirical model developed by \citet{2010ApJ...714L.202T}, \citet{2013ApJ...768L..37T} and \cite{2015ApJ...813...21M} to include scatter in galaxy luminosities and predict the evolution of the Ultraviolet Luminosity Function (UVLF) for $z>5$ when there is stochasticity present. We model galaxies with a conditional luminosity function (CLF) approach, which includes a log-normal scatter parameter $\Sigma$, in conjunction with a \citet{2001MNRAS.321..372J} halo mass function (HMF). In general, we expect stochasticity to play a prominent role for the rarest objects and show this in Figure~\ref{fig:SBS}, where a higher value of scatter implies the formation of an over-luminous galaxy hosted inside a lower mass halo. This process is primarily facilitated by the exponential shape at the massive end of the HMF. We show that this leads to a broader exponential end for the UVLF which can be interpreted as a departure of the Schechter shape for high values of $\Sigma$.

A lower bound for the parameter $\Sigma$ can be estimated through the assembly distribution of the host dark matter halos. Modeling from \citet{2018ApJ...856...81R} infers $\Sigma \sim 0.2$ for $z > 2$ as the lower limit. To maintain consistency with the observed UVLFs, a flattening in the median galaxy luminosity versus halo mass relation $L_{c}(M_{h})$ is required past a characteristic point. From this, further evolution in the UVLF can progress under these possible scenarios:

\begin{enumerate}
  \item The critical point for flattening in $L_{c}(M_{h})$ is a luminosity threshold, i.e. the median galaxy luminosity is capped at some critical luminosity. Physically, this could represent a scenario where feedback is triggered by a process that depends on the halo growth rate, i.e. quasar-mode AGN feedback. In this scenario the LF retains the steep exponential slope typical of a Schechter shape under redshift evolution for $z > 6$. 
  \item The critical point for flattening in $L_{c}(M_{h})$ is a mass threshold, i.e. the median galaxy luminosity flattens for halo masses greater than the critical mass. This could correspond to a scenario where star formation regulated by a process dependent on the halo mass, such as radio-mode AGN feedback. Here, the LF departs from the Schechter profile at higher redshifts as fewer halos can form with a mass exceeding the critical mass threshold.
\end{enumerate}

The predicted UVLFs for $6 < z < 12$ shown in Figure~\ref{fig:LFs0} highlight the differences between both scenarios. The critical mass scenario (solid colored lines) deviates further from Schechter (black line) in redshift, while the critical luminosity case (dashed colored lines) is closer to the Schechter shape. Additionally, in the cases for $\Sigma \lesssim 0.2$, the predicted evolved UVLF can match the scatter free ($\Sigma = 0$) LFs derived from former models almost completely. We note that there is degeneracy for $\Sigma > 0.2$ between the critical luminosity and critical mass scenarios as there is a natural point where a flattening cannot accommodate the broadening of the UVLF \footnote{Both of these scenarios can also be independent and occur concurrently}. With this, in Figure~\ref{fig:Sc5} we show that this over-broadening can help provide a tentative estimate for the upper limit $\Sigma \lesssim 0.25(0.3)$ assuming a critical mass (critical luminosity) scenario. Generally, $\Sigma>0$ changes the functional shape for the UVLF at $z \gtrsim 6$, explaining the observational results indicating a possible departure from Schechter LF reported by \citet{2014MNRAS.440.2810B} and \citet{Ono2017}.

Often, we consider properties from rare extreme objects as key indicators for unraveling the cosmic evolution of structure. In this work, we see that $\Sigma$ encodes the significance of stochasticity in the formation of these extreme objects. Large area surveys have the potential to discriminate between values of $\Sigma$. For this, we also cast predictions for future missions such as WFIRST for the number of objects that could be detected per square degree. Even with the minimal case scatter $\Sigma = 0.2$, we predict over an order of magnitude increase in $m_{AB} < 23.5$ objects found for $z > 8$ compared to $\Sigma = 0$ models. This number can be boosted an additional order of magnitude for $\Sigma = 0.4$. However, discriminating between critical luminosity or critical mass scenarios becomes challenging because of degeneracies in predictions for when $\Sigma > 0.2$. Thus a more robust approach in constraining $\Sigma$ would be to supplement the former with measurements of the clustering strength around bright objects, which is a complementary probe of $\Sigma$ \citep{2018ApJ...856...81R} enabling a joint constraint on $\Sigma$ and the dominant scenario of $L_{c}(M_{h})$ evolution at high $z$. The latter method can be readily enabled with the upcoming launch of the James Webb Space Telescope in 2021, thus the prospect of utilising both approaches in measuring $\Sigma$ would be able to provide valuable insight into fundamental processes into the mechanisms that drive star formation during the Epoch of Reionization.

{\acknowledgements{
We thank the referee for their valuable input to this manuscript. In addition, we thank P. Behroozi and A. Melatos for their helpful discussions on methods of deconvolution. This research was conducted by the Australian Research Council Centre of Excellence for All Sky Astrophysics in 3 Dimensions (ASTRO 3D), through project number CE170100013. K.R is additionally supported through the Research Training Program Scholarship from the Australian Government. C.M. acknowledges support through the NASA Hubble Fellowship grant HST-HF2-51413.001-A awarded by the Space Telescope Science Institute, which is operated by the Association of Universities for Research in Astronomy, Inc., for NASA, under contract NAS5-26555.}}

\bibliography{scatterpaper}

\begin{thebibliography}{}
\expandafter\ifx\csname natexlab\endcsname\relax\def\natexlab#1{#1}\fi

\bibitem[{{Allen} {et~al.}(2018){Allen}, {Behroozi}, \&
  {Ma}}]{2018arXiv181205733A}
{Allen}, M., {Behroozi}, P., \& {Ma}, C.-P. 2018, arXiv e-prints,
  arXiv:1812.05733

\bibitem[{{Barone-Nugent} {et~al.}(2015){Barone-Nugent}, {Wyithe}, {Trenti},
  {Treu}, {Oesch}, {Bouwens}, {Illingworth}, \&
  {Schmidt}}]{2015MNRAS.450.1224B}
{Barone-Nugent}, R.~L., {Wyithe}, J.~S.~B., {Trenti}, M., {et~al.} 2015,
  \mnras, 450, 1224

\bibitem[{Behroozi {et~al.}(2010)Behroozi, Conroy, \& Wechsler}]{Behroozi2010}
Behroozi, P.~S., Conroy, C., \& Wechsler, R.~H. 2010, The Astrophysical
  Journal, 717, 379

\bibitem[{Behroozi \& Silk(2015)}]{Behroozi2015}
Behroozi, P.~S., \& Silk, J. 2015, The Astrophysical Journal, 799, 32

\bibitem[{Behroozi {et~al.}(2013)Behroozi, Wechsler, \& Conroy}]{Behroozi2013}
Behroozi, P.~S., Wechsler, R.~H., \& Conroy, C. 2013, The Astrophysical
  Journal, 770, 57

\bibitem[{{Bernard} {et~al.}(2016){Bernard}, {Carrasco}, {Trenti}, {Oesch},
  {Wu}, {Bradley}, {Schmidt}, {Bouwens}, {Calvi}, {Mason}, {Stiavelli}, \&
  {Treu}}]{2016ApJ...827...76B}
{Bernard}, S.~R., {Carrasco}, D., {Trenti}, M., {et~al.} 2016, \apj, 827, 76

\bibitem[{{Bhowmick} {et~al.}(2018){Bhowmick}, {Campbell}, {Di Matteo}, \&
  {Feng}}]{2018MNRAS.480.3177B}
{Bhowmick}, A.~K., {Campbell}, D., {Di Matteo}, T., \& {Feng}, Y. 2018, \mnras,
  480, 3177

\bibitem[{Bouwens {et~al.}(2010)Bouwens, Illingworth, Oesch, Stiavelli, van
  Dokkum, Trenti, Magee, Labb{\'{e}}, Franx, Carollo, \&
  Gonzalez}]{Bouwens2010}
Bouwens, R.~J., Illingworth, G.~D., Oesch, P.~A., {et~al.} 2010, The
  Astrophysical Journal, 709, L133

\bibitem[{{Bouwens} {et~al.}(2015){Bouwens}, {Illingworth}, {Oesch}, {Trenti},
  {Labb{\'e}}, {Bradley}, {Carollo}, {van Dokkum}, {Gonzalez}, {Holwerda},
  {Franx}, {Spitler}, {Smit}, \& {Magee}}]{2015ApJ...803...34B}
{Bouwens}, R.~J., {Illingworth}, G.~D., {Oesch}, P.~A., {et~al.} 2015, \apj,
  803, 34

\bibitem[{Bowler {et~al.}(2016)Bowler, Dunlop, McLure, \& McLeod}]{Bowler2016}
Bowler, R. A.~A., Dunlop, J.~S., McLure, R.~J., \& McLeod, D.~J. 2016, Monthly
  Notices of the Royal Astronomical Society, 466, 3612

\bibitem[{Bowler {et~al.}(2014)Bowler, Dunlop, McLure, Rogers, McCracken,
  Milvang-Jensen, Furusawa, Fynbo, Taniguchi, Afonso, Bremer, \&
  F{\`{e}}vre}]{2014MNRAS.440.2810B}
Bowler, R. A.~A., Dunlop, J.~S., McLure, R.~J., {et~al.} 2014, Monthly Notices
  of the Royal Astronomical Society, 440, 2810

\bibitem[{Bowler {et~al.}(2015)Bowler, Dunlop, McLure, McCracken,
  Milvang-Jensen, Furusawa, Taniguchi, F{\`{e}}vre, Fynbo, Jarvis, \&
  Häu{\ss}ler}]{Bowler2015}
---. 2015, Monthly Notices of the Royal Astronomical Society, 452, 1817

\bibitem[{Bradley {et~al.}(2012)Bradley, Trenti, Oesch, Stiavelli, Treu,
  Bouwens, Shull, Holwerda, \& Pirzkal}]{Bradley2012}
Bradley, L.~D., Trenti, M., Oesch, P.~A., {et~al.} 2012, The Astrophysical
  Journal, 760, 108

\bibitem[{{Bruzual} \& {Charlot}(2003)}]{2003BRUZUAL&CHARLOT}
{Bruzual}, G., \& {Charlot}, S. 2003, \mnras, 344, 1000

\bibitem[{{Cooray} \& {Milosavljevi{\'c}}(2005)}]{2005ApJ...627L..89C}
{Cooray}, A., \& {Milosavljevi{\'c}}, M. 2005, \apjl, 627, L89

\bibitem[{{Croton} {et~al.}(2006){Croton}, {Springel}, {White}, {De Lucia},
  {Frenk}, {Gao}, {Jenkins}, {Kauffmann}, {Navarro}, \&
  {Yoshida}}]{2006MNRAS.365...11C}
{Croton}, D.~J., {Springel}, V., {White}, S.~D.~M., {et~al.} 2006, \mnras, 365,
  11

\bibitem[{{Gardner} {et~al.}(2006){Gardner}, {Mather}, {Clampin}, {Doyon},
  {Greenhouse}, {Hammel}, {Hutchings}, {Jakobsen}, {Lilly}, {Long}, {Lunine},
  {McCaughrean}, {Mountain}, {Nella}, {Rieke}, {Rieke}, {Rix}, {Smith},
  {Sonneborn}, {Stiavelli}, {Stockman}, {Windhorst}, \&
  {Wright}}]{2006SSRv..123..485G}
{Gardner}, J.~P., {Mather}, J.~C., {Clampin}, M., {et~al.} 2006, \ssr, 123, 485

\bibitem[{{Grogin} {et~al.}(2011){Grogin}, {Kocevski}, {Faber}, {Ferguson},
  {Koekemoer}, {Riess}, {Acquaviva}, {Alexander}, {Almaini}, {Ashby}, {Barden},
  {Bell}, {Bournaud}, {Brown}, {Caputi}, {Casertano}, {Cassata}, {Castellano},
  {Challis}, {Chary}, {Cheung}, {Cirasuolo}, {Conselice}, {Roshan Cooray},
  {Croton}, {Daddi}, {Dahlen}, {Dav{\'e}}, {de Mello}, {Dekel}, {Dickinson},
  {Dolch}, {Donley}, {Dunlop}, {Dutton}, {Elbaz}, {Fazio}, {Filippenko},
  {Finkelstein}, {Fontana}, {Gardner}, {Garnavich}, {Gawiser}, {Giavalisco},
  {Grazian}, {Guo}, {Hathi}, {H{\"a}ussler}, {Hopkins}, {Huang}, {Huang},
  {Jha}, {Kartaltepe}, {Kirshner}, {Koo}, {Lai}, {Lee}, {Li}, {Lotz}, {Lucas},
  {Madau}, {McCarthy}, {McGrath}, {McIntosh}, {McLure}, {Mobasher},
  {Moustakas}, {Mozena}, {Nandra}, {Newman}, {Niemi}, {Noeske}, {Papovich},
  {Pentericci}, {Pope}, {Primack}, {Rajan}, {Ravindranath}, {Reddy}, {Renzini},
  {Rix}, {Robaina}, {Rodney}, {Rosario}, {Rosati}, {Salimbeni}, {Scarlata},
  {Siana}, {Simard}, {Smidt}, {Somerville}, {Spinrad}, {Straughn}, {Strolger},
  {Telford}, {Teplitz}, {Trump}, {van der Wel}, {Villforth}, {Wechsler},
  {Weiner}, {Wiklind}, {Wild}, {Wilson}, {Wuyts}, {Yan}, \&
  {Yun}}]{2011ApJS..197...35G}
{Grogin}, N.~A., {Kocevski}, D.~D., {Faber}, S.~M., {et~al.} 2011, \apjs, 197,
  35

\bibitem[{Ishigaki {et~al.}(2018)Ishigaki, Kawamata, Ouchi, Oguri, Shimasaku,
  \& Ono}]{Ishigaki2018}
Ishigaki, M., Kawamata, R., Ouchi, M., {et~al.} 2018, The Astrophysical
  Journal, 854, 73

\bibitem[{{Jenkins} {et~al.}(2001){Jenkins}, {Frenk}, {White}, {Colberg},
  {Cole}, {Evrard}, {Couchman}, \& {Yoshida}}]{2001MNRAS.321..372J}
{Jenkins}, A., {Frenk}, C.~S., {White}, S.~D.~M., {et~al.} 2001, \mnras, 321,
  372

\bibitem[{{Komatsu} {et~al.}(2011){Komatsu}, {Smith}, {Dunkley}, {Bennett},
  {Gold}, {Hinshaw}, {Jarosik}, {Larson}, {Nolta}, {Page}, {Spergel},
  {Halpern}, {Hill}, {Kogut}, {Limon}, {Meyer}, {Odegard}, {Tucker}, {Weiland},
  {Wollack}, \& {Wright}}]{2011ApJS..192...18K}
{Komatsu}, E., {Smith}, K.~M., {Dunkley}, J., {et~al.} 2011, \apjs, 192, 18

\bibitem[{{Livermore} {et~al.}(2017){Livermore}, {Finkelstein}, \&
  {Lotz}}]{2017ApJ...835..113L}
{Livermore}, R.~C., {Finkelstein}, S.~L., \& {Lotz}, J.~M. 2017, \apj, 835, 113

\bibitem[{{Lotz} {et~al.}(2017){Lotz}, {Koekemoer}, {Coe}, {Grogin}, {Capak},
  {Mack}, {Anderson}, {Avila}, {Barker}, {Borncamp}, {Brammer}, {Durbin},
  {Gunning}, {Hilbert}, {Jenkner}, {Khandrika}, {Levay}, {Lucas}, {MacKenty},
  {Ogaz}, {Porterfield}, {Reid}, {Robberto}, {Royle}, {Smith},
  {Storrie-Lombardi}, {Sunnquist}, {Surace}, {Taylor}, {Williams}, {Bullock},
  {Dickinson}, {Finkelstein}, {Natarajan}, {Richard}, {Robertson}, {Tumlinson},
  {Zitrin}, {Flanagan}, {Sembach}, {Soifer}, \&
  {Mountain}}]{2017ApJ...837...97L}
{Lotz}, J.~M., {Koekemoer}, A., {Coe}, D., {et~al.} 2017, \apj, 837, 97

\bibitem[{{Madau} {et~al.}(1996){Madau}, {Ferguson}, {Dickinson}, {Giavalisco},
  {Steidel}, \& {Fruchter}}]{1996madau}
{Madau}, P., {Ferguson}, H.~C., {Dickinson}, M.~E., {et~al.} 1996, \mnras, 283,
  1388

\bibitem[{Mashian {et~al.}(2015)Mashian, Oesch, \& Loeb}]{Mashian2015}
Mashian, N., Oesch, P.~A., \& Loeb, A. 2015, Monthly Notices of the Royal
  Astronomical Society, 455, 2101

\bibitem[{{Mason} {et~al.}(2015{\natexlab{a}}){Mason}, {Trenti}, \&
  {Treu}}]{2015ApJ...813...21M}
{Mason}, C.~A., {Trenti}, M., \& {Treu}, T. 2015{\natexlab{a}}, \apj, 813, 21

\bibitem[{{Mason} {et~al.}(2015{\natexlab{b}}){Mason}, {Treu}, {Schmidt},
  {Collett}, {Trenti}, {Marshall}, {Barone-Nugent}, {Bradley}, {Stiavelli}, \&
  {Wyithe}}]{2015ApJ...805...79M}
{Mason}, C.~A., {Treu}, T., {Schmidt}, K.~B., {et~al.} 2015{\natexlab{b}},
  \apj, 805, 79

\bibitem[{{Morishita} {et~al.}(2018){Morishita}, {Trenti}, {Stiavelli},
  {Bradley}, {Coe}, {Oesch}, {Mason}, {Bridge}, {Holwerda}, {Livermore},
  {Salmon}, {Schmidt}, {Shull}, \& {Treu}}]{Morishita2018}
{Morishita}, T., {Trenti}, M., {Stiavelli}, M., {et~al.} 2018, ArXiv e-prints,
  arXiv:1809.07604

\bibitem[{Moster {et~al.}(2010)Moster, Somerville, Maulbetsch, van~den Bosch,
  Macci{\`{o}}, Naab, \& Oser}]{Moster2010}
Moster, B.~P., Somerville, R.~S., Maulbetsch, C., {et~al.} 2010, The
  Astrophysical Journal, 710, 903

\bibitem[{Mu{\~{n}}oz \& Loeb(2011)}]{Munoz2011}
Mu{\~{n}}oz, J.~A., \& Loeb, A. 2011, The Astrophysical Journal, 729, 99

\bibitem[{{Mutch} {et~al.}(2016){Mutch}, {Geil}, {Poole}, {Angel}, {Duffy},
  {Mesinger}, \& {Wyithe}}]{2016MNRAS.462..250M}
{Mutch}, S.~J., {Geil}, P.~M., {Poole}, G.~B., {et~al.} 2016, \mnras, 462, 250

\bibitem[{{Oesch} {et~al.}(2015){Oesch}, {van Dokkum}, {Illingworth},
  {Bouwens}, {Momcheva}, {Holden}, {Roberts-Borsani}, {Smit}, {Franx},
  {Labb{\'e}}, {Gonz{\'a}lez}, \& {Magee}}]{2015ApJ...804L..30O}
{Oesch}, P.~A., {van Dokkum}, P.~G., {Illingworth}, G.~D., {et~al.} 2015,
  \apjl, 804, L30

\bibitem[{{Oesch} {et~al.}(2016){Oesch}, {Brammer}, {van Dokkum},
  {Illingworth}, {Bouwens}, {Labb{\'e}}, {Franx}, {Momcheva}, {Ashby}, {Fazio},
  {Gonzalez}, {Holden}, {Magee}, {Skelton}, {Smit}, {Spitler}, {Trenti}, \&
  {Willner}}]{2016ApJ...819..129O}
{Oesch}, P.~A., {Brammer}, G., {van Dokkum}, P.~G., {et~al.} 2016, \apj, 819,
  129

\bibitem[{Oesch {et~al.}(2016)Oesch, Brammer, van Dokkum, Illingworth, Bouwens,
  Labb{\'{e}}, Franx, Momcheva, Ashby, Fazio, Gonzalez, Holden, Magee, Skelton,
  Smit, Spitler, Trenti, \& Willner}]{Oesch2016}
Oesch, P.~A., Brammer, G., van Dokkum, P.~G., {et~al.} 2016, The Astrophysical
  Journal, 819, 129

\bibitem[{{Oke} \& {Gunn}(1983)}]{1983ApJ...266..713O}
{Oke}, J.~B., \& {Gunn}, J.~E. 1983, \apj, 266, 713

\bibitem[{Ono {et~al.}(2017)Ono, Ouchi, Harikane, Toshikawa, Rauch, Yuma,
  Sawicki, Shibuya, Shimasaku, Oguri, Willott, Akhlaghi, Akiyama, Coupon,
  Kashikawa, Komiyama, Konno, Lin, Matsuoka, Miyazaki, Nagao, Nakajima,
  Silverman, Tanaka, Taniguchi, \& Wang}]{Ono2017}
Ono, Y., Ouchi, M., Harikane, Y., {et~al.} 2017, Publications of the
  Astronomical Society of Japan, 70, doi:10.1093/pasj/psx103

\bibitem[{Reddick {et~al.}(2013)Reddick, Wechsler, Tinker, \&
  Behroozi}]{Reddick2013}
Reddick, R.~M., Wechsler, R.~H., Tinker, J.~L., \& Behroozi, P.~S. 2013, \apj,
  771, 30

\bibitem[{{Ren} {et~al.}(2018){Ren}, {Trenti}, \&
  {Mutch}}]{2018ApJ...856...81R}
{Ren}, K., {Trenti}, M., \& {Mutch}, S.~J. 2018, \apj, 856, 81

\bibitem[{Salmon {et~al.}(2018)Salmon, Coe, Bradley, Brada{\v{c}}, Strait,
  Paterno-Mahler, Huang, Oesch, Zitrin, Acebron, Cibirka, Kikuchihara, Oguri,
  Brammer, Sharon, Trenti, Avila, Ogaz, Andrade-Santos, Carrasco, Cerny,
  Dawson, Frye, Hoag, Jones, Mainali, Ouchi, Rodney, Stark, \&
  Umetsu}]{Salmon2018}
Salmon, B., Coe, D., Bradley, L., {et~al.} 2018, The Astrophysical Journal,
  864, L22

\bibitem[{Schmidt {et~al.}(2014)Schmidt, Treu, Trenti, Bradley, Kelly, Oesch,
  Holwerda, Shull, \& Stiavelli}]{Schmidt2014}
Schmidt, K.~B., Treu, T., Trenti, M., {et~al.} 2014, The Astrophysical Journal,
  786, 57

\bibitem[{{Spergel} {et~al.}(2015){Spergel}, {Gehrels}, {Baltay}, {Bennett},
  {Breckinridge}, {Donahue}, {Dressler}, {Gaudi}, {Greene}, {Guyon}, {Hirata},
  {Kalirai}, {Kasdin}, {Macintosh}, {Moos}, {Perlmutter}, {Postman},
  {Rauscher}, {Rhodes}, {Wang}, {Weinberg}, {Benford}, {Hudson}, {Jeong},
  {Mellier}, {Traub}, {Yamada}, {Capak}, {Colbert}, {Masters}, {Penny},
  {Savransky}, {Stern}, {Zimmerman}, {Barry}, {Bartusek}, {Carpenter}, {Cheng},
  {Content}, {Dekens}, {Demers}, {Grady}, {Jackson}, {Kuan}, {Kruk}, {Melton},
  {Nemati}, {Parvin}, {Poberezhskiy}, {Peddie}, {Ruffa}, {Wallace}, {Whipple},
  {Wollack}, \& {Zhao}}]{2015arXiv150303757S}
{Spergel}, D., {Gehrels}, N., {Baltay}, C., {et~al.} 2015, ArXiv e-prints,
  arXiv:1503.03757

\bibitem[{Stanek {et~al.}(2006)Stanek, Evrard, Bohringer, Schuecker, \&
  Nord}]{Stanek2006}
Stanek, R., Evrard, A.~E., Bohringer, H., Schuecker, P., \& Nord, B. 2006, The
  Astrophysical Journal, 648, 956

\bibitem[{Stefanon {et~al.}(2017)Stefanon, Labb{\'{e}}, Bouwens, Brammer,
  Oesch, Franx, Fynbo, Milvang-Jensen, Muzzin, Illingworth, F{\`{e}}vre,
  Caputi, Holwerda, McCracken, Smit, \& Magee}]{Stefanon2017}
Stefanon, M., Labb{\'{e}}, I., Bouwens, R.~J., {et~al.} 2017, The Astrophysical
  Journal, 851, 43

\bibitem[{Sun \& Furlanetto(2016)}]{Sun2016}
Sun, G., \& Furlanetto, S.~R. 2016, Monthly Notices of the Royal Astronomical
  Society, 460, 417

\bibitem[{{Tacchella} {et~al.}(2013){Tacchella}, {Trenti}, \&
  {Carollo}}]{2013ApJ...768L..37T}
{Tacchella}, S., {Trenti}, M., \& {Carollo}, C.~M. 2013, \apjl, 768, L37

\bibitem[{{Trenti} {et~al.}(2015){Trenti}, {Perna}, \&
  {Jimenez}}]{2015ApJ...802..103T}
{Trenti}, M., {Perna}, R., \& {Jimenez}, R. 2015, \apj, 802, 103

\bibitem[{{Trenti} {et~al.}(2010){Trenti}, {Stiavelli}, {Bouwens}, {Oesch},
  {Shull}, {Illingworth}, {Bradley}, \& {Carollo}}]{2010ApJ...714L.202T}
{Trenti}, M., {Stiavelli}, M., {Bouwens}, R.~J., {et~al.} 2010, \apjl, 714,
  L202

\bibitem[{{Trenti} {et~al.}(2011){Trenti}, {Bradley}, {Stiavelli}, {Oesch},
  {Treu}, {Bouwens}, {Shull}, {MacKenty}, {Carollo}, \&
  {Illingworth}}]{2011ApJ...727L..39T}
{Trenti}, M., {Bradley}, L.~D., {Stiavelli}, M., {et~al.} 2011, \apjl, 727, L39

\bibitem[{Vale \& Ostriker(2004)}]{Vale2004}
Vale, A., \& Ostriker, J.~P. 2004, Monthly Notices of the Royal Astronomical
  Society, 353, 189

\bibitem[{Vale \& Ostriker(2008)}]{Vale2008}
---. 2008, Monthly Notices of the Royal Astronomical Society, 383, 355

\bibitem[{{Wyithe} {et~al.}(2011){Wyithe}, {Yan}, {Windhorst}, \&
  {Mao}}]{Wyithe2011}
{Wyithe}, J.~S.~B., {Yan}, H., {Windhorst}, R.~A., \& {Mao}, S. 2011, \nat,
  469, 181

\bibitem[{{Yang} {et~al.}(2005){Yang}, {Mo}, {Jing}, \& {van den
  Bosch}}]{2005MNRAS.358..217Y}
{Yang}, X., {Mo}, H.~J., {Jing}, Y.~P., \& {van den Bosch}, F.~C. 2005, mnras,
  358, 217

\bibitem[{Yang {et~al.}(2003)Yang, Mo, \& van~den Bosch}]{Yang2003}
Yang, X., Mo, H.~J., \& van~den Bosch, F.~C. 2003, Monthly Notices of the Royal
  Astronomical Society, 339, 1057

\bibitem[{{Zitrin} {et~al.}(2014){Zitrin}, {Zheng}, {Broadhurst}, {Moustakas},
  {Lam}, {Shu}, {Huang}, {Diego}, {Ford}, {Lim}, {Bauer}, {Infante}, {Kelson},
  \& {Molino}}]{2014ApJ...793L..12Z}
{Zitrin}, A., {Zheng}, W., {Broadhurst}, T., {et~al.} 2014, \apjl, 793, L12

\bibitem[{{Zitrin} {et~al.}(2015){Zitrin}, {Labb{\'e}}, {Belli}, {Bouwens},
  {Ellis}, {Roberts-Borsani}, {Stark}, {Oesch}, \&
  {Smit}}]{2015ApJ...810L..12Z}
{Zitrin}, A., {Labb{\'e}}, I., {Belli}, S., {et~al.} 2015, \apjl, 810, L12

\end{thebibliography}
\bibliographystyle{aasjournal}

\appendix
\section{Comparison of iterative deconvolution method} \label{apdx:a}

\begin{figure}[ht!]
	\centerline{\includegraphics[angle=-00, scale=0.70]{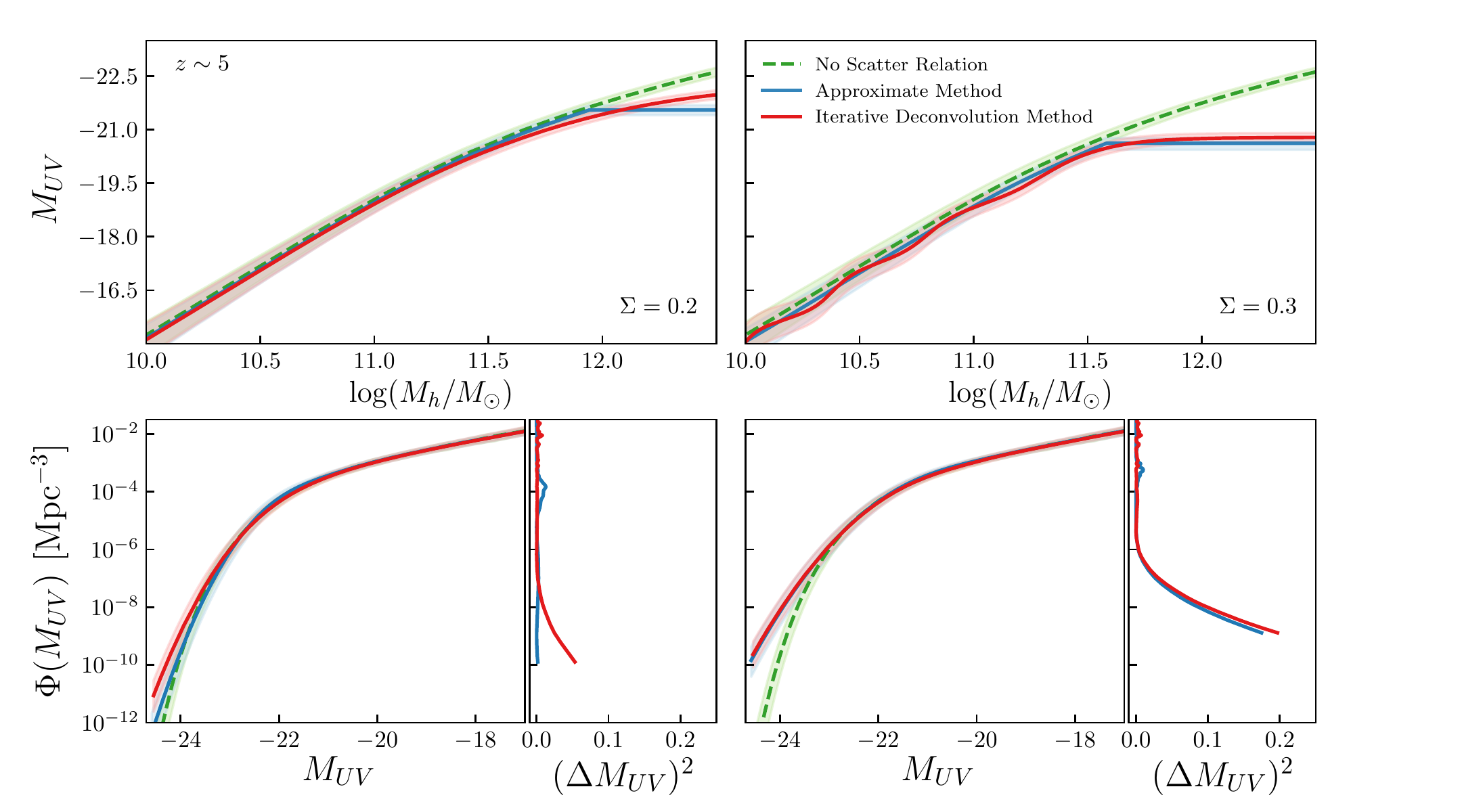}}
	\caption{\small Comparing the $z \sim 5$ modeled LFs using two methods: (1) An iterative deconvolution method as described in \citet{Behroozi2010} (solid red). (2) An approximate method involving a least squares minimization after a scaling and flattening (solid blue). The $\Sigma = 0$ cases are represented by the dashed green line. We show the output assuming a $\Sigma = 0.2$ (left panels) and $\Sigma = 0.3$ (right panels). The left panels are the respective median galaxy luminosity versus halo mass relations when we account for scatter (solid lines). The corresponding LFs modeled using Equation~\ref{eqn:lf}. The squared residuals in $M_{UV}$ are also shown. The shaded regions are the $1 \sigma$ uncertainties.}
	\label{fig:comparison}
\end{figure}

The deconvolution method detailed explicitly in \citet{2018arXiv181205733A} can be used to determine the median stellar mass versus halo mass relation after assuming some $\Sigma$ can adapted to calculate $L_{c}$. The method can be summarized by the following steps:

\begin{enumerate}
  \item We perform a forward calculation of the modeled LF, $\phi_{M}(\log L')$ after assuming some scatter by using Equation~\ref{eqn:lf}. The initial guess of $L_{c}' = L_{c}(M_{h}, \Sigma = 0)$  is obtained by direct abundance matching of the calibration UVLF with the \citet{2001MNRAS.321..372J} HMF.
  \item We abundance match our modeled LF, $\phi_{M}(\log L')$ with the calibration UVLF. This will indicate how our modeled $\log L'$ must transform to reobtain the correct calibration LF.
  \item We shift the median galaxy luminosity versus halo mass relation, $L_{c}'$ according to the transformation required in $\log L'$. By doing so, we change our the distribution of $\Sigma$.
  \item Steps $(1)$ to $(3)$ is repeated using our transformed $L_{c}'$ relation.
  \item We terminate loop when the squared residual of our magnitudes at $\Phi(M_{UV} = -23)$ falls under a specified tolerance. This magnitude is a fixed point that corresponds to the brightest observation point used in our calibration LF. The tolerances used in Figure~\ref{fig:comparison} are $\varepsilon = 0.0006 (0.006)$ for $\Sigma = 0.2 (0.3)$.
\end{enumerate}

In Figure~\ref{fig:comparison}, we see that both methods recover adequate solutions that describe well the calibration UVLF up to the current bright-end observational limit of $M_{UV} \sim -23$. The iterative deconvolution method slightly fits the faint end better, but struggles to converge onto the bright end as seen by the presence of oscillatory behaviour in the median galaxy luminosity versus halo mass relation. This exhibited behaviour is purely numerical and is a byproduct of pushing the tolerance $\varepsilon$ to arbitrarily high levels. This is simply because the calibration LF cannot be deconvolved exactly with our defined scatter distribution. This is a limitation of using direct abundance matching as a deconvolution method when the information of our initial conditions, either our calibration LF or the assumption of lognormal stochasticity, is imperfect.

From this, we see that the usage of either relations is insignificant to the overall study as the flattening of the bright end are observed in both methods. Primarily this suggests that a large contribution to the bright end of the LF is due to over-luminous galaxies within smaller halos instead of average galaxies inside extremely massive halos.

\end{document}